\documentclass[showpacs,amsmath,amssymb,twocolumn,floatfix,prl]{revtex4}
\usepackage[dvips]{graphicx}
\input{epsf}

\begin{document}

\title{Nonequilibrium stationary states with ratchet effect}


\author{G.~Cristadoro$^{(a),(b)}$ and 
D.L.~Shepelyansky$^{(b)}$}
\homepage[]{http://www.lpt.irsamc.ups-tlse.fr/~dima}
\affiliation{$^{(a)}$    Center for Nonlinear and Complex Systems, 
Dipartimento di Scienze Chimiche, Fisiche e Matematiche, 
Universit\`a dell'Insubria, Via Valleggio 11,  and
Istituto Nazionale di Fisica della Materia, 
Unit\`a di Como, 22100 Como, Italy\\
$^{(b)}$Laboratoire de Physique Th\'eorique, 
UMR 5152 du CNRS, Universit\'e P. Sabatier, 31062 Toulouse Cedex 4, France
}

\date{October 20, 2004}

\begin{abstract} 
An ensemble of particles in thermal equilibrium
at temperature $T$, modeled  by Nos\`e-Hoover dynamics,
moves on a triangular  lattice of oriented semi-disk
elastic scatterers. Despite the scatterer asymmetry
a directed transport is clearly ruled out by 
the second law of thermodynamics. 
Introduction of a polarized zero mean monochromatic field creates 
a directed stationary flow with nontrivial 
dependence on temperature and field parameters.
We give a theoretical estimate of directed current induced
by a microwave field in an antidot superlattice
in semiconductor  heterostructures.
\end{abstract}


\maketitle
According to  the second law of thermodynamics
there is no stationary directed transport 
in spatially periodic asymmetric systems  in   
thermal equilibrium \cite{smol,feynman}.
However, a time periodic parameter variation
may drive such a system out of equilibrium 
leading to the emergence of stationary transport
whose direction depends  non trivially
on parameters. Such directed transport appears in
systems with  noise, fluctuations and dissipation and
is now called  Brownian motor or ratchet (see e.g. reviews
\cite{belinicher,hanggi,reimann}). The ratchet effect
has a generic nature and it has been observed 
in various physical systems including
semiconductor heterostructures \cite{linke},
cold atoms in a laser field \cite{grynberg},
vortices in superconductors  \cite{mooij,nori,ustinov}
and macroporous silicon membranes under pressure
oscillations \cite{muller}. It has also  important
applications in biological systems as discussed
in \cite{hanggi,prost}.
 
In spite of a great recent interest to ratchets 
the theoretical research is mainly
concentrated on one-dimensional models
(see e.g. \cite{reimann}). Also, since the ratchet 
behavior is usually rather complex,
an overdamped limit is used very often 
to obtain analytical parameter dependence
even if in this regime a directed transport
is absent for {\it ac} zero mean force  \cite{reimann}.
To understand in a better way the global properties
of ratchets and their dependence on such important physical
parameters as temperature $T$ and  driving strength ${\bf f}$,
we analyze here a generic case when
{\it ac}-driving affects a Maxwell thermostat ensemble
of noninteracting particles moving 
in an asymmetric two-dimensional (2D) periodic structure.
This structure is composed of triangular 2D-lattice
of rigid semi-disks of radius $r_d$ as shown in Fig.~1 (insert).
The distance $R$ between disk centers is fixed to be
$R=2 r_d$ and we assume that collisions with semi-disks are
elastic. Free particle motion between semi-disks is affected by
a polarized monochromatic force 
${\bf f} = f (\cos \theta, \sin \theta) \cos \omega t$
with frequency $\omega$, strength $f$ and polarization
angle $\theta$ to $x$-axis. It is also assumed that
particles are in thermal equilibrium  and 
at $f=0$ their velocities are
given by the Maxwell distribution at temperature $T$.
Fig.~1 shows that in this system {\it ac}-force generates
stationary  directed transport.
In numerical computations we put $r_d$, particle mass $m$, unit of time and
Boltzmann's constant $k$ to be equal to unity.
\vglue -0.3cm 
\begin{figure}[th!]  
\centerline{\epsfxsize=8.5cm\epsffile{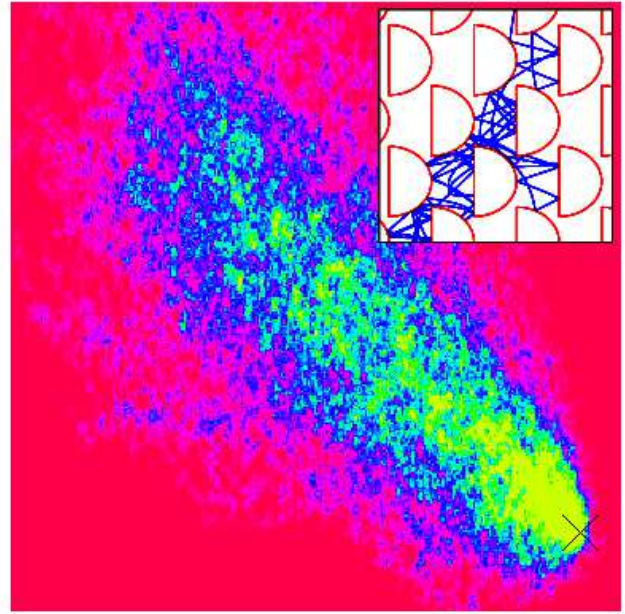}}
\vglue -0.3cm 
\caption{(color online) 
Density distribution averaged over the time interval
$0 \leq t \leq 5 \cdot 10^5$ and obtained from dynamics of
200 particles given by the Nos\`e-Hoover equations 
at thermostat temperature $T=24$.
The region of distribution is  $x=[-2050,150]$, $y=[-300,1900]$.
Initially particles are placed at $x=y=0$ (cross)
with random velocities. 
Density is proportional to color changing from zero (red/black) 
to maximum (yellow/white). The parameters of driving force
are $f=16$,  $\omega=1.5$ and $\theta=\pi/8$.
The relaxation time scale of the thermostat is $\tau=\sqrt{50}$.
Insert shows one trajectory on small scale moving between semi-disks.
}
\label{fig1}       
\end{figure}

In order to put particles in thermal equilibrium we choose the elegant 
method of the Nos\`e-Hoover thermostat 
(see e.g. \cite{hover,hover1,klages} and Refs. therein). In this method
the motion of a particle is affected by an effective friction $\gamma$
which keeps the average kinetic energy 
$\langle \mathbf{p}^2/2 \rangle$ equal to a given thermostat
temperature $T$. In this way the dynamics of particle is described
by the  equations:
\begin{equation}
\label{enh}
\mathbf{\dot q} = \mathbf{p}/m \; , \;\; 
\mathbf{\dot p} = \mathbf{F} - \gamma \mathbf{p} \; ,  \;\;
\dot{\gamma} = [\mathbf{p}^2/(2mT) - 1]/\tau^2 
\end{equation}
where $\mathbf{q}, \mathbf{p}$ are particle coordinate and momentum,
$\mathbf{F}$ is a sum of {\it ac}-force and force of elastic collisions
with semi-disks, and $\tau$ is the time scale of relaxation to equilibrium. 
\begin{figure}[t!]
\centerline{\epsfxsize=8.5cm\epsffile{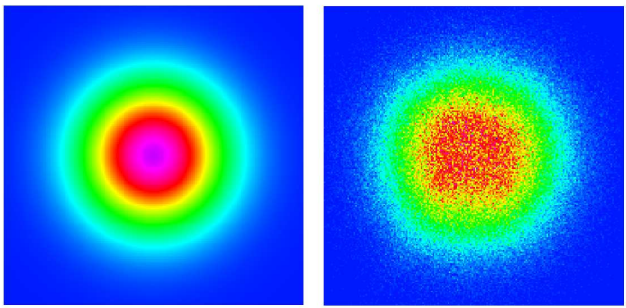}}
\vglue -0.2cm
\caption{(color online) 
Steady state distribution in 2D momentum plane $(p_x,p_y)$, density
is proportional to color changing from zero (blue/black)
to maximum (rose-violet/gray). Left: the Maxwell  distribution
at temperature of Fig.~1; right: distribution obtained numerically
from the Nos\`e-Hoover thermostat for the case of Fig.~1.
}
\label{fig2}
\end{figure}

It is known that the Nos\`e-Hoover thermostat works well only if
the dynamics is sufficiently chaotic \cite{hover,hover1,klages}. 
In some cases, e.g. for the Galton board, the Nos\`e-Hoover
thermostat gives noticeable deviations from 
the Maxwell distribution \cite{klages}.
To check that in our case this
method really gives a thermal equilibrium we analyze 
the steady state distribution in the momentum space obtained by numerical
Runge-Kutta integration of Eqs.(\ref{enh}). 
Our results show that a small {\it ac}-force 
is needed to make chaotic dynamics between 
semi-disks more homogeneous
and to produce a stable Maxwell thermal equilibrium 
which is not sensitive to variation of relaxation rate $1/\tau$
(Fig.3, insert). 
At large force the numerical data
show that 2D steady state in the momentum space is still close to the
Maxwell distribution (Fig.~2) even if
the {\it ac}-driving produces a clear ratchet effect shown in Fig.~1.
The dependence of steady state on temperature  closely follows
the Maxwell distribution in
 momentum space $p=| \mathbf{p} |$ as shown in Fig.~3.
Thus we may conclude that the dynamics
given by the Nos\`e-Hoover equations allows efficiently
investigate the effects of {\it ac}-driving on 
particles in thermal equilibrium. The numerical data 
show that this driving generates 
 a strong ratchet effect (Fig.~1) 
with directed transport wich  
depends on temperature and parameters of driving force.
\begin{figure}[t!]  
\centerline{\epsfxsize=8.5cm\epsffile{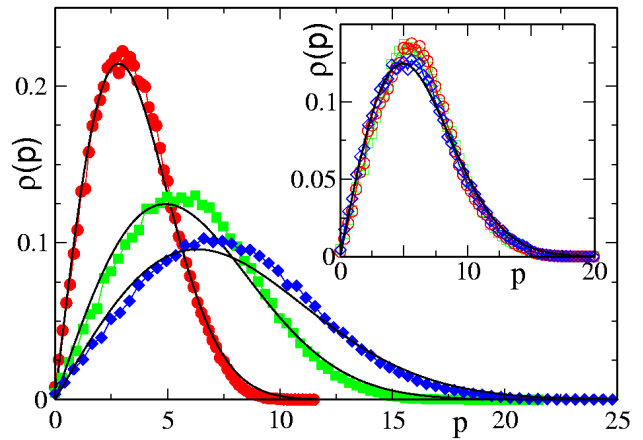}}
\vglue -0.2cm 
\caption{(color online) 
Thermal distribution $\rho$ in momentum $p=| \mathbf{p} |$ for
different values of temperature $T=8, 16, 40$
(from narrow to broad distribution, respectively,
shown by curves and symbols). For each temperature value
the curve gives the Maxwell distribution  and the symbols
show the numerical data for the Nos\`e-Hoover thermostat
($\tau=\sqrt{50}$)
in presence of {\it ac}-driving with parameters of Fig.~1.
Insert shows the stability of Nos\`e-Hoover thermostat
at small {\it ac}-force ($f=0.5$, $\omega=1.5$, $\theta=\pi/8$)
with respect to variation of  relaxation time 
$\tau^2 = 1$ (circles), 50 (squares),
100 (diamonds) at fixed temperature $T=24$; 
curve shows the Maxwell distribution. All numerical data 
are obtained from one long trajectory with $t \leq 5 \cdot 10^5$.
}
\label{fig3}       
\end{figure}

To understand the properties of this directed transport
we first analyze the dependence of averaged friction
$\langle \gamma \rangle$ on driving strength $f$ and
temperature $T$. The value of $\langle \gamma \rangle$
is obtained by averaging over long time interval
during numerical integration of Eqs.(\ref{enh}) for one trajectory.
We also checked that averaging over a few trajectories gives 
the same result. The data are shown in Fig.~4.
They are well described by a global scaling given by 
\begin{equation}
\label{gamma}
\langle \gamma \rangle = C r_d m^{-1/2} f^2/T^{3/2} \; .
\end{equation}
Small deviations seen at low $T $ appear because
of strong driving force which starts to
modify significantly the particle velocity 
distribution in this regime.  
The numerical constant $C$ is only weakly dependent on $\theta$
and $\omega$ changing by  50\% to 30\% when $\theta$ 
changes from $0$ to $\pi/2$ and $\omega$  changes by a factor 10,
respectively. The dependence (\ref{gamma}) clearly tells that 
in presence of driving force the thermostat creates an effective
friction force $\mathbf{f}_f = - \langle \gamma \rangle \mathbf{p}$ acting
on particle propagation with an effective friction
constant $\langle \gamma \rangle$.
Surprisingly, this friction coefficient varies with $f$ and $T$
according to Eq.(\ref{gamma}) but in a large range remains independent
of the relaxation time $\tau$ (Fig.~4 insert). 
We note that the particle dynamics
in absence of thermostat but in presence of  friction force 
$\mathbf{f}_f = - \gamma \mathbf{p}$ with constant friction coefficient
$\gamma$ has been analyzed in \cite{chep} where it was shown that
{\it ac}-force generates a directed transport on semi-disk lattice.

To understand the origin of the dependence (\ref{gamma})
we put forward the following heuristic arguments.
The driving force gives a diffusive  energy growth
during a dissipative time scale $1/\gamma$
so that
\begin{equation}
\label{diff}
(\Delta E)^2 \sim D_E /\gamma \; , \;\; D_E \sim f^2 v l  \; ,
\end{equation}
where the diffusion rate in energy is $D_E \sim {\dot{E}}^2 \tau_c 
\sim f^2 v l$ 
and the mean-free path $l \sim R \sim r_d \sim 1$ 
determines the collision time $\tau_c = l/v$. In the Maxwell
equilibrium the particle velocity is $v \sim (T/m)^{1/2}$
and the fact that the driving force does not modify
the velocity distribution  implies that $\Delta E \sim T$
so that the diffusive growth is stopped
by effective friction $\gamma  \sim D_E/T^2
\sim  r_d m^{-1/2}f^2/T^{3/2}$ in agreement with (\ref{gamma}).
In fact there is a close relation to results \cite{chep}
where the thermostat is absent but a friction force
$\mathbf{f}_f = - \gamma \mathbf{p}$ with constant $\gamma$
affects particle dynamics. In that case {\it ac}-driving force
heats a particle up to energy $E \sim (r_d f^2/m^{1/2}\gamma)^{2/3}$
while in presence of thermostat the energy is fixed
by temperature $T \sim E$ that imposes a convergence
to the stationary state with effective friction 
given by (\ref{gamma}) \cite{note}. 
\begin{figure}[t!]
\epsfxsize=3.2in
\epsffile{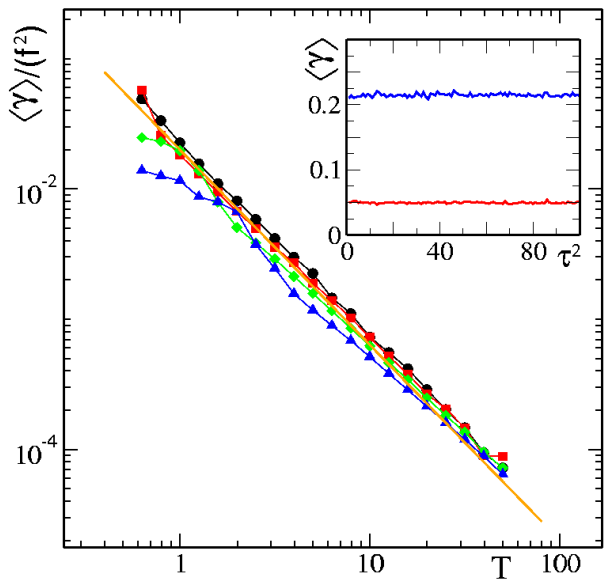}
\vglue -0.3cm
\caption{(color online) 
Dependence of rescaled average friction coefficient 
$\langle \gamma \rangle /f^2$ on temperature $T$
for $f=$4 (circles), 8 (squares), 16 (diamonds) 32 
(triangles) (top to bottom) at fixed
$\omega =1.5$, $\theta=0$, $\tau^2=50$.
The straight line shows dependence (\ref{gamma})
with $C=0.02$.
Insert shows that $\langle \gamma \rangle$ is robust
against variation of $\tau^2$ in the interval [1,100]
(with step 1);
data are shown for $f=16$, $\omega =1.5$, $\theta=0$,
$T=$ 8 (top curve) and 24 (bottom curve).
}
\label{fig4}       
\end{figure}

The dependence of average velocity $v_f$ of the ratchet flow
on $\langle \gamma \rangle$ for various values
of driving strength $f$  is shown in Fig.~5.
Globally, the flow velocity $v_f$ grows with increase 
of $\gamma$. Two regimes are clearly seen:
$v_f \approx r_d\langle \gamma \rangle$ for
$\langle \gamma \rangle  < \gamma_c$ and
$v_f \approx (r_d^2 f \langle \gamma \rangle /m)^{1/3}/10$ for
$\langle \gamma \rangle  > \gamma_c \approx f^{1/2}/[30(r_d m)^{1/2}]$.
In fact this dependence is very close to the one found
in a model with fixed $\gamma$ \cite{chep}. 
As a result, from (\ref{gamma})
we obtain the dependence of 
flow velocity on temperature:
\begin{eqnarray}
\label{flow}
v_f /v & \approx & r_d f /50T \;\;\;,\; (T<T_c); \\
\label{flow1}
v_f /v & \approx & (r_d f/8T)^2  \;,\; (T>T_c);
\end{eqnarray}
where $v=(2T/m)^{1/2}$ is the thermal velocity
and $T_c \approx r_d f$ is linked to $\gamma_c$
obtained from Fig.~5. The transition between
two regimes takes place when the energy given 
by {\it ac}-force to particle between two collisions
becomes larger than thermal energy ($T<T_c$).  
In that case the effect of driving is strong
and $v_f \sim f \tau_c/m \sim r_d f/(mT)^{1/2}$
leading to (\ref{flow}).
For  $T>T_c$ thermal fluctuations are
strong and the ratchet effect appears only in the
second order of force $f$ giving (\ref{flow1}). 
The numerical factors in (\ref{flow}),(\ref{flow1})
are taken for the case $\theta=0$ from Figs.~4,5.
We note that Eqs. (\ref{gamma})-(\ref{flow1})
are derived in the regime of relatively weak
friction $\langle \gamma \rangle \ll \omega$
and relaxation rate $1/\tau \ll \omega$.
Another important point is that the dependence (\ref{flow}),(\ref{flow1})
is robust with respect to variation of scatter geometry,
e.g. introduction of additional disk scatterer in the center of unit cell
eliminates all  collisionless paths  but gives no significant modifications
(see Fig.~5).
\begin{figure}[t!]
\epsfxsize=3.2in
\epsfysize=3.2in
\epsffile{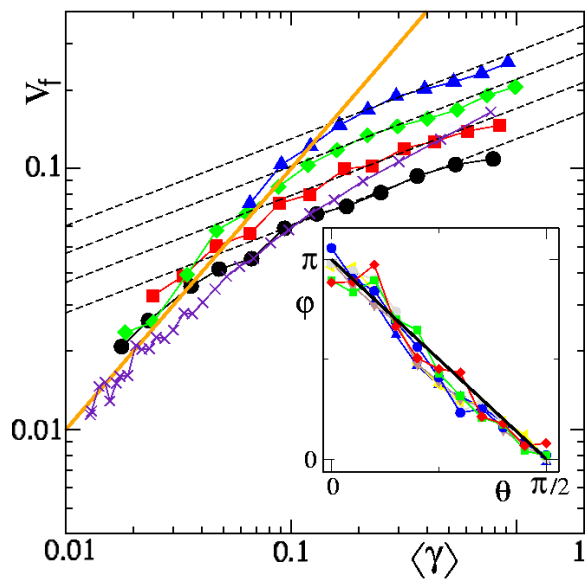}
\vglue -0.3cm
\caption{(color online) 
Dependence of the absolute value of average velocity of particle
flow $v_f$ on average friction $\langle \gamma \rangle$
for the parameters of Fig.~4 with 
$f=4$, 8, 16, 32 (same symbols, from bottom to top).
Dashed lines show the scaling dependence 
$v_f \sim (f \langle \gamma \rangle)^{1/3}$
at large $\langle \gamma \rangle$ for different $f$ values;
the full line shows scaling $v_f = \langle \gamma \rangle$
at small $\langle \gamma \rangle$. Crosses show data for the same
parameters as for squares ($f=8$) but with additional circular
scatterer added in the center of unit cell to eliminate
orbits with a straight flight through the whole system
(see text). Insert shows the dependence of 
flow direction angle $\varphi$ on polarization angle $\theta$;
data are given for $f=8$, $\omega=1.5$, $\tau^2=50$
and $4 \leq T \leq 11$; full line shows average dependence
$\varphi=\pi-2\theta$.
}
\label{fig5}       
\end{figure}

The dependence of flow directionality, determined through
angle $ \varphi$ 
($\mathbf{v_f}=v_f (\cos \varphi, \sin  \varphi)$),
on the polarization of {\it ac}-force is shown in Fig.~5 (insert).
In average, it is satisfactory described by the relation
$\varphi = \pi - 2\theta$ 
(similar dependence was seen in \cite{chep}). 
On a qualitative ground,
we may say that at $\theta=0$ due to friction
a particle becomes trapped between semi-disks of a unit cell
that gives a directed transport to the left
while for $\theta=\pi/2$ vertical oscillations push particle
to the right in presence of friction. 
The linear dependence $\varphi = \pi - 2\theta$ interpolates between
these two limits.
However, a more quantitative derivation is needed.

It is interesting to apply the approach
developed above to other types of thermostat. It is possible
to realize the semi-disk Galton board with antidot superlattices
for 2D electron gas in semiconductor heterostructures.
With such structures
the Galton board of disks has already  been implemented (see e.g.
\cite{weiss}) and effects of microwave radiation has been 
studied \cite{kvon1}. For disk antidots like those in
\cite{weiss,kvon1} the ratchet effect is absent due to symmetry of
antidot. However, for semi-disk antidot lattice
strong ratchet effect should appear. To find its properties
we should take into account that in this case we have
the Fermi-Dirac thermostat with the Fermi energy $E_F \gg T$.
Due to that in (\ref{diff}) the particle velocity $v$ is equal
to the Fermi velocity $v_F=(2E_F/m)^{1/2}$ independent of $T$.
This modification gives the average friction $\gamma_F$ for the Fermi gas
\begin{equation}
\label{fgamma}
\gamma_F = C f^2 v_F r_d/T^2 \approx v_f/r_d\; ,
\end{equation}
where we  kept the same
numerical constant $C \sim 1/50$.
In fact (\ref{fgamma}) follows from $D_E \sim f^2 v_F r_d$
(see (\ref{diff}))  and $\gamma_F \sim D_E/T^2$.
The second equality in (\ref{fgamma}) appears due to
the fact that $E_F \gg T_c$ implying the regime (\ref{flow1})
with $v_f \sim \gamma_F r_d$. 
Of course, only a small fraction $T/E_F$
of electrons near $E_F$ contributes to this ratchet flow.
Hence, the current $I$ per one semi-disk row is
\begin{equation}
\label{current}
I \sim e r_d n_e v_f T/E_F \sim C e r_d^3 \sqrt{n_e} f^2 / ( T \hbar) \; .
\end{equation}
where we used that for the  2D electron Fermi gas
$E_F = \pi n_e \hbar^2/m$.
We note that in semiconductor antidot lattices
like in \cite{weiss,kvon1} the effective 
mass $m$ is about 15 times smaller compared to the electron mass.
For a typical parameters of semi-disk antidot
lattice with electron density $n_e \sim 10^{12} cm^{-2}$,
$r_d \sim 1\mu m$,
field strength per electron charge $f/e \sim 1 V/cm  $
and $T \sim 10 K$ 
we obtain $v_f/v_F \sim 10^{-4}$. 
At these parameters
$E_F \sim 150 K$, $v_F \sim 3 \cdot 10^7 cm/sec$
and the current $I \sim 10^{-9} A$ 
is sufficiently large
to be observed experimentally. The result (\ref{current})
is based on the semiclassical estimate for the diffusion rate
$D_E$ which assumes that the energy of microwave photon
is larger than the level spacing $\Delta$ inside
one unit cell: 
$\hbar \omega > \Delta \approx 2\pi \hbar^2/(m r_d^2)$.
In the opposite limit $\hbar \omega \ll \Delta$,
{\it ac}-driving is in the quantum adiabatic regime
when the excitation in energy is very weak.
Thus for $r_d \sim 1 \mu m$ we have
$\Delta \approx 5 \cdot 10^{-6} eV \approx 0.05 K$   
and the directed transport
appears only for $\omega/2\pi >  1 GHz$.
In experiments \cite{linke}
the frequency  was deeply in the adiabatic
regime with $\omega/2\pi \sim 100Hz$
and the directed transport was absent at zero mean force.
We also note that in the quantum case the ratchet transport 
should disappear as soon as the amplitude of oscillations 
$f/m \omega^2$ induced by  {\it ac}-force becomes smaller
 than the wavelenght $\hbar/m v_F$ at the Fermi level. 
Thus the ratchet survives only for $\omega<\sqrt{f v_F/\hbar}$. 
For field strenght of $1 V/cm$ this gives an approximate  bound at $30 GHz$.  

In summary, we showed that zero mean {\it ac}-force
applied to particles being in thermal equilibrium
in asymmetric periodic potential creates
a directed transport flow. Its direction is 
efficiently changed by polarization of the force.
We also established the dependence of the flow
velocity $v_f$ on temperature and driving field
strength for the Maxwell 
[Eqs. (\ref{flow}),(\ref{flow1}),(\ref{fgamma})]
and the Fermi-Dirac [Eqs. (\ref{fgamma}),(\ref{current})]
thermostats.

We thank Alexei Chepelianskii, Kvon Ze Don 
and Sergey Vitkalov for useful discussions.


\begin{thebibliography}{99}
\bibitem{note} We note that a numerical
  constant  $C \approx 0.02$ in (\ref{gamma}) (see Fig.~4) 
  corresponds to average dependence $v^2=2T=(f^2/\gamma)^{2/3}/7$
  from Fig.~5 in \cite{chep}.
\end{thebibliography}
\end{document}